\documentclass[a4paper,twoside,10pt]{letter}
\usepackage{graphicx,saj,multicol,subeqnarray}

\newcommand{\HII}{H\,{\sc ii}}

\def\arcmin{\hbox{$^\prime$}}
\def\arcsec{\hbox{$^{\prime\prime}$}}

\def\p0{\phantom{0}}

\def\papertype{}

\def\udc{...}
\begin{document}
\baselineskip=3.1truemm
\columnsep=.5truecm
\newenvironment{lefteqnarray}{\arraycolsep=0pt\begin{eqnarray}}
{\end{eqnarray}\protect\aftergroup\ignorespaces}
\newenvironment{lefteqnarray*}{\arraycolsep=0pt\begin{eqnarray*}}
{\end{eqnarray*}\protect\aftergroup\ignorespaces}
\newenvironment{leftsubeqnarray}{\arraycolsep=0pt\begin{subeqnarray}}
{\end{subeqnarray}\protect\aftergroup\ignorespaces}
%


\markboth{\eightrm NEW 20-CM RADIO-CONTINUUM STUDY OF THE SMALL MAGELLANIC CLOUD: PART I - IMAGES}
{\eightrm G. F. WONG, et. al.}

{\ }

\publ

\type

{\ }


\title{New 20-cm Radio-continuum study of the Small Magellanic Cloud: part I - images}


\authors{G. F. Wong$^{1}$, M. D. Filipovi\'c$^{1}$, E. J. Crawford$^{1}$ A. De Horta$^{1}$, T. Galvin$^{1}$, }
\authors{D. Dra\v skovi\'c$^{1}$, J. L. Payne$^{1}$ }

\vskip3mm


\address{$^1$University of Western Sydney, Locked Bag 1797, Penrith South DC, NSW 1797, Australia}
\Email{m.filipovic}{uws.edu.au}


\dates{April XX, 2011}{April XX, 2011}


\summary{We present and discuss new high-sensitivity and resolution radio-continuum images of the Small Magellanic Cloud (SMC) at $\lambda$=20~cm ($\nu$=1.4~GHz). The new images were created by merging 20-cm radio-continuum  archival data, from the Australian Telescope Compact Array  and  the Parkes radio-telescope. Our images span from $\sim$10\arcsec\ to $\sim$150\arcsec\ in resolution and sensitivity of r.m.s.$\geq$0.5~mJy/beam. These images will be used in future studies of the SMC's intrinsic sources and its overall extended structure.   }



\begin{multicols}{2}
{

\section{1. INTRODUCTION}

The well established proximity of the Small Magellanic Cloud (SMC; $\sim$60~kpc; Hilditch et al.~2005) and its location in one of the coldest areas of the radio sky towards South Celestial Pole, allows observation of its radio emissions to be made without the interference of our own galaxy's dust, gas and stars (Galactic Foreground Radiation). This means that, not only the study of its intrinsic properties including its extended emission and polarisation are of great interest, but that it is an ideal location to study other objects such as supernova remnants (SNRs; Filipovi{\' c} et al. 2005, 2008), \HII\ regions and Planetary Nebulae (PNe; Filipovi{\' c} et al. 2009a) which are difficult to study in our own galaxy and other galaxies that are further away.  

Over the last 40 years extensive radio-continuum surveys of the Small Magellanic Cloud (SMC) have been made including, interferometric observations made using the Molonglo Synesis Telescope (MOST; Ye et al. 1995) and Australian Telescope Compact Array (ATCA; Filipovi{\' c} et al. 2002, Payne et al. 2004, Filipovi{\' c} et al. 2009b, Mao et al. 2008, Dickel et al. 2010), and single dish observations from the 64-m Parkes radio-telescope (Filipovi{\' c} et al. 1997, 1998). Most of these surveys, however, suffer from either low resolution, poor sensitivity and/or poor {\it uv}-coverage. 

In this paper, we present and discuss a method of merging various radio-continuum observations of the SMC at $\lambda$=20~cm ($\nu$=1.4~GHz) in an attempt to make the best use of the currently available data prior to the next generation of radio-telescope surveys (such as ASKAP, MEERCAT and SKA). By combining a large amount of existing observational data of the SMC and with the latest generation of computer power we can create various new high-resolution and high-sensitivity images. 

The newly constructed images are analysed and the difference between the various SMC images created at 20~cm are discussed. In \S2 we describe the observational data and reduction techniques. In \S3 we present our new maps with a brief discussion and in\S4 is the conclusion. The astrophysical interpretation of sources found in the surveys will be presented in subsequent papers.

\section{2. OBSERVATIONAL DATA}

\subsection{2.1 Observational Data Details}

To create the highest fidelity and resolution image of the SMC at 20~cm to date, we looked for mosaic observations that covered the SMC region. After an extensive search on the Australian Telescope Online Archive\footnote{http://atoa.atnf.csiro.au} (ATOA), we retrieved three projects C159, C1197 and C1288 that observed the whole SMC.

All these three projects taken from the ATOA were ATCA mosaic observations of the SMC. Used as part of a survey of neutral hydrogen emission in the SMC by Staveley-Smith et al. (1997; see paper for observation details), project C159 (Fig.~1)provided the first large scale image using a mosaic mode. The ATCA project C1197 (Fig.~3) contained two sets of observations using different array configurations (H214 and H75). Observations of the H75 array configurations were part of a study into the cool gas in the Magellanic Stream (see Matthews et al. 2009 for observation details). The ATCA project C1288 (Fig.~2) was used as part of a study of magnetic fields of the SMC, carrying out radio Faraday rotation and optical starlight polarisation data (see Mao et al. (2008) for observation details). The Parkes data was obtained from a radio-continuum study (Fig.~4) of the SMC (Filipovi{\' c} et al. 1997). The source 1934-638 was used for the primary calibration and the source 0252-712 as the secondary calibrator for all ATCA images. A brief summary of the three ATCA projects can be found in Table~1.

}
\end{multicols}

\centerline{{\bf Table 1.} ATOA mosaic observational data used in imaging of SMC.}
\vskip1mm
\centerline{
\begin{tabular}{lllccc}
\hline
\emph{ATCA Project}&\emph{Date Observed}& \emph{Array}&\emph{Bandwidth(MHz)}\\
\hline
C159 & 06-10-1992 to 07-10-1992 & 375&4\\
& 09-10-1992 to 14-10-1992 & 375&4 \\
C1288 & 10-07-2004 to 18-07-2004 & 6A&128\\
C1197& 20-10-2003 to 01-11-2003  &H214&128\\
& 31-07-2005 to 02-08-2005  & H75&128\\
& 09-09-2006 to 10-09-2006  & H75&128\\
& 12-09-2006 to 15-09-2006  & H75&128\\
& 19-09-2006 to 22-09-2006  & H75&128\\
\hline
\end{tabular}}
\vskip.5cm
\begin{multicols}{2}
{

\subsection{2.2 Image Creation}

The software packages \textsc{miriad} (Sault and Killeen 2010) and \textsc{karma} (Gooch 2006) were used for the data reduction and analysis. Initial high-resolution images were produced from the full dataset using the \textsc{miriad} multi-frequency synthesis (Sault and Wieringa 1994) with natural weighting. 

The joint-deconvolution method (Cornwell 1988) was used in the imaging process to give better {\it uv}-coverage from the different pointings that were overlapping each other.

The deconvolution process employed \textsc{miriad} tasks \textsc{mossdi} and \textsc{mosmem} (or combination of both). Ideal for the deconvolution of point source emissions, \textsc{mossdi} is an SDI variance of the clean algorithm that is designed for Mosaics (Steer et al. 1984). The map created from ATCA project C1288 (Fig.~2), consisted mostly of point sources due to the long baselines, therefore \textsc{mossdi} was used to deconvolve it.

While on the other hand, \textsc{mosmem} is a method that uses maximum entropy (Cornwell 1989) and is ideal for extended objects. This task was used for ATCA projects C159 (Fig.~1) and C1197 (Fig.~3) where the short baseline were used and extended emission dominates.

The various resolution maps were created by restricting the radial distance in the {\it uv}-plane of the three ATCA projects. The combined maps with various resolutions (with the exception of the highest resolution map), were created using a combination of both the \textsc{miriad} task \textsc{mossdi} and \textsc{mosmem}. This method struck the balance between the two deconvolution tasks and its effects on point source emission and extended emission. Using \textsc{mossdi} to clean the mosaic with a low number of iterations till the side lobes of strong point sources are no longer present, removed any interference. Then using \textsc{mosmem} to complete the deconvolution process allowed the extended emissions from the data of the short baselines to remain.  

To create the best possible SMC mosaic image at 20~cm, we examined the radial distance in the {\it uv}-plane of all three ATCA projects combined (Fig.~5), and identified where significant gaps between baselines existed. The first significant gap was found at 3k$\lambda$ (see this image in Fig.~6). Using this as a starting point we were able to process and successfully create an image. Subsequently we increased the radial distance from 3k$\lambda$ to 6k$\lambda$ (Fig.~7) and later 16k$\lambda$ (Fig.~8).

We also merged all mosaics with the Parkes observations to provide the essential and important short spacing data (Stanimirovi\'c 2002).

The new images presented in this paper shows the individual maps (from all ATCA projects; Table~2) and combined maps that are merged with Parkes data. Table~3. lists the details of the various combined images.

}
\end{multicols}

\centerline{{\bf Table 2.}The details of ATCA projects of SMC mosaics at 20-cm.}
\vskip1mm
\centerline{
\begin{tabular}{lccccc}
\hline
\emph{ATCA}&\emph{Beam Size}&\emph{PA} &\emph{r.m.s.}\\
\emph{Project}& (arcsec) &(degree)& (mJy/beam)\\
\hline
C159  & 98.0$\times$98.0    &\p00 & 1.5\\
C1288 & 17.8$\times$12.2    &26   & 0.7\\
C1197 & 158.2$\times$124.8  &85   & 1.0\\
\hline
\end{tabular}}
\vskip.5cm

\centerline{{\bf Table 3.}The details of three SMC merged mosaics at 20-cm.}
\vskip1mm
\centerline{
\begin{tabular}{ccccc}
\hline
\emph{uvdistance}&\emph{Beam Size}&\emph{PA}&\emph{r.m.s.}\\
(k$\lambda$)& (arcsec) &(degree)& (mJy/beam)\\
\hline
\p03  &  48.4 $\times$ 33.6   & 26 & 0.5\\
\p06  &  30.1 $\times$ 20.1   & 27 & 0.5\\
16    &  16.3 $\times$ \p09.5 & 22 & 0.5\\
\hline
\end{tabular}}
\vskip.5cm

\begin{multicols}{2}
{

\section{3. RESULTS AND DISCUSSION}

In Figs. 1 to 3, we show the individual intensity mosaic maps of the SMC from ATCA projects C159, C1288 and C1197 respectively and in Fig.~4 we reproduce the Parkes (single-dish) image from Filipovi{\' c} et al. (1997). The Figs.~6, 7 and 8 are images at {\it uv} distance of 3, 6 and 16 k$\lambda$ respectively (Table~3).

All these images can be downloaded from the Astronomy Digital Image Library (ADIL) at http://adil.ncsa.uiuc.edu/.

\subsection{3.1 Individual SMC Mosaics at $\lambda$=20~cm}

Comparing the individual images of the SMC at 20~cm created from the ATCA projects, we can see the effects of the different array configurations. While shorter baseline images such as C159 (Fig.~1) and C1197 (Fig.~3) shows much intensive extended emission, compared to C1288 image (Fig.~2; 6A array) which contains point sources at higher resolution.  

Fig.~1 was imaged from the C159 observations and later merged with the data from Parkes single-dish radio-telescope. It contains a combination of extended emission and point source emission. From the individual images (Figs.~1 to 4), Fig.~1 has a balanced combination of the different types of emission and resolution.  

Using an array configuration of 6A, image in Fig.~2 (ATCA project C1288) has the longest baseline and only point sources can be seen. However, Fig.~2 has the highest resolution compared to the other individual images presented in this paper and dominated by the radio point source population. 

Created from ATCA project C1197, in Fig.~3 the most prominent feature is extended emission. Although this image suffer from poor resolution due to the dominance of shorter baselines, which upon closer examination means a lack of detail of individual objects. 

In Fig. 4 we show observation taken from single dish telescope (Parkes). As a result, this image has the lowest resolution compared to Figs.~1 to 3 which are observations taken from an array interferometer. The main characteristic of Fig.~4 is the extended emission and the shape of the SMC galaxy.

We note that the area coverage of these three mosaics is different. Therefore, when combining the mosaics we only used the individual observations that overlapped with each other.

\subsection{3.2 New Combined SMC Mosaics at $\lambda$=20~cm}

Figs.~6-8 are combination images with various resolutions, the features of the image have a combination of point source emission and extended emission.   

We also point that some differences between various images can be attributed to slightly different deconvolution techniques and careful flagging of highly noisy observational data. Our new high-resolution and high-sensitive analysis of these analysis will be presented in future papers.

}
\end{multicols}

\centerline{\includegraphics[width=1\textwidth,angle=-90]{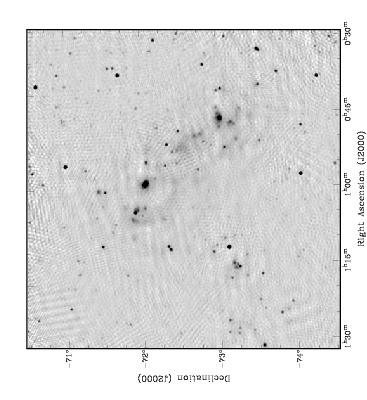}}
\figurecaption{1.}{ATCA project C159 radio-continuum total intensity image of the SMC  merged with Parkes. The synthesised beam is 98\arcsec\ and the r.m.s=1.5~mJy/beam. }

\centerline{\includegraphics[width=0.9\textwidth,angle=-90]{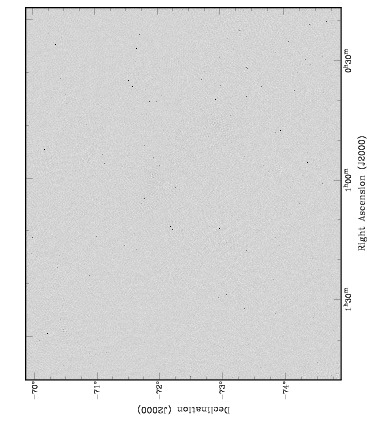}}
\figurecaption{2.}{ATCA Project C1288 radio-continuum total intensity image of the SMC merged with Parkes. The synthesised beam is 17.8\arcsec$\times$12.2\arcsec\ and the r.m.s=0.7~mJy/beam. }

\centerline{\includegraphics[width=0.85\textwidth,angle=-90]{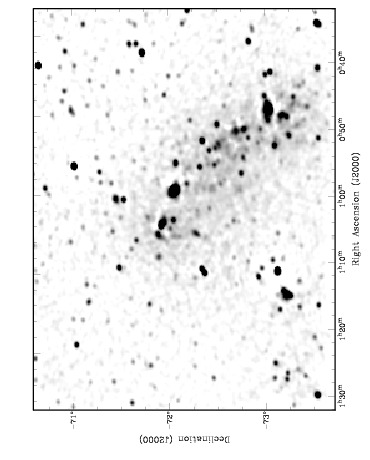}}
\figurecaption{3.}{ATCA Project C1197 radio-continuum total intensity image of the SMC. The synthesised beam is 158.2\arcsec$\times$124.8\arcsec\ and the r.m.s=1.0~mJy/beam. }

\centerline{\includegraphics[width=1\textwidth,angle=-90]{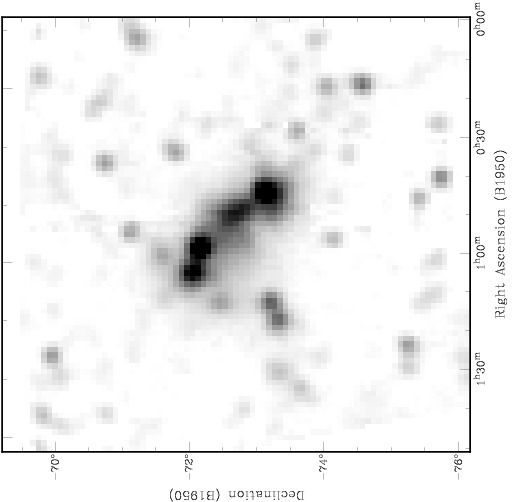}}
\figurecaption{4.}{A radio-continuum total intensity of SMC taken from Parkes retrieved from Filipovi{\' c} et al. (1997). The synthesised beam is 13.8\arcmin\ and r.m.s=15~mJy/beam.}

\centerline{\includegraphics[width=0.7\textwidth,angle=-90]{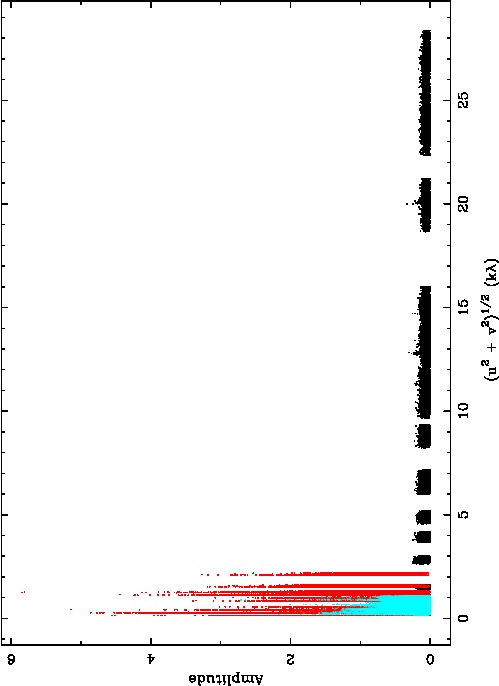}}
\figurecaption{5.}{Amplitude vs. {\it uv}-distance graph of the three projects overlapping with each other. Blue colour represents observational data from ATCA project C1197, red from C159 and black from C1288. }

\centerline{\includegraphics[width=0.78\textwidth,angle=-90]{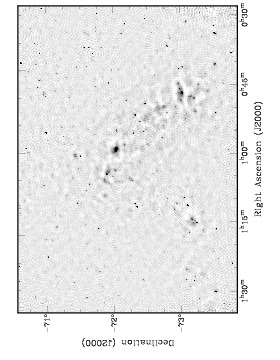}}
\figurecaption{6.}{A radio-continuum total intensity images of the SMC at 3k$\lambda$. The synthesised beam is 48.4\arcsec$\times$33.6\arcsec\ and the r.m.s=0.5~mJy/beam. }

\centerline{\includegraphics[width=0.82\textwidth,angle=-90]{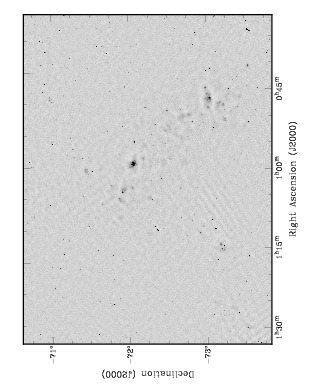}}
\figurecaption{7.}{A radio-continuum total intensity images of the SMC at 6k$\lambda$. The synthesised beam is 30.1\arcsec$\times$20.1\arcsec\ and the r.m.s=0.5~mJy/beam. }

\centerline{\includegraphics[width=0.72\textwidth,angle=-90]{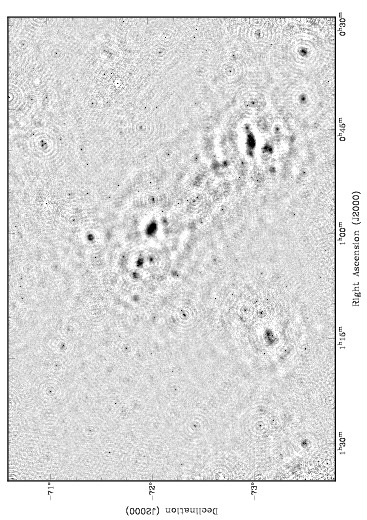}}
\figurecaption{8.}{A radio-continuum total intensity images of the SMC at 16k$\lambda$. The synthesised beam is 16.3\arcsec$\times$9.5\arcsec\ and the r.m.s=0.5~mJy/beam. }

\begin{multicols}{2}
{

\section{4. CONCLUSION}

In this paper we present and discuss new high-sensitivity and resolution radio-continuum images of the SMC at 20~cm. The new images were created from merging sensitive 20-cm mosaic radio surveys, from the ATOA, and later the data from Parkes 64-m radio-telescope. These images will be used in future studies of the SMC's intrinsic sources and overall structure.


\acknowledgements{The Australia Telescope Compact Array and Parkes radio-telescope is part of the Australia Telescope National Facility which is funded by the Commonwealth of Australia for operation as a National Facility managed by CSIRO. This paper includes archived data obtained through the Australia Telescope Online Archive (http://atoa.atnf.csiro.au). }


\references

Cornwell, T.J.: 1988, \journal{Astron. Astrophys.}, \vol{202}, 316.

Cornwell, T.J.: 1989, \journal{Highlights of Astronomy}, \vol{8}, 547.

Dickel, J.R.; Gruendl, R.A.; McIntyre, V.J., Shaun W.A.: 2010, \journal{Astron. J.}, \vol{140}, 1511.

Filipovi{\'c}, M.D., Jones, P.A., White, G.L, Haynes, R.F, Klein, U., Wielebinski, R.: 1997, \journal{Astron. Astrophys. Suppl. Series}, \vol{121}, 321.

Filipovi{\'c}, M.D., Haynes, R.F., White, G.L., Jones, P.A.: 1998, \journal{Astron. Astrophys. Suppl. Series}, \vol{130}, 421.

Filipovi{\'c}, M.D., Bohlsen, T., Reid, W, Staveley-Smith, L., Jones, P.A, Nohejl, K., Goldstein, G.: 2002, \journal{Mon. Not. R. Astron. Soc.}, \vol{335}, 1085.

Filipovi{\'c}, M.D., Payne, J.L., Reid, W., Danforth, C.W., Staveley-Smith, L., Jones, P.A., White, G.L.: 2005, \journal{Mon. Not. R. Astron. Soc.}, \vol{364}, 217.

Filipovi{\'c}, M.D., Haberl, F., Winkler, P.F., Pietsch, W., Payne, J.L., Crawford, E.J., de Horta, A.Y., Stootman, F.H., Reaser, B.E.: 2008, \journal{Astron. Astrophys.}, \vol{485}, 63.

Filipovi{\'c}, M.D., Cohen, M., Reid, W.A., Payne, J.L., Parker, Q.A., Crawford, E.J., Boji\v ci\'c, I.S., de Horta, A.Y., Hughes, A., Dickel, J., Stootman, F.: 2009a, \journal{Mon. Not. R. Astron. Soc.}, \vol{399}, 769.

Filipovi{\'c}, M.D., Crawford E.~J., Hughes A., Leverenz H., de Horta A.~Y., Payne J.~L., Staveley-Smith L., Dickel J.~R., Stootman F.~H., White G.~L.: 2009b, in van Loon J.~T., Oliveira J.~M., eds, \journal{IAU Symposium Vol. 256 of IAU Symposium}, PDF8

Gooch, R.: 2008, \textsc{Karma} Users Manual, ATNF, Sydney.

Hilditch, R.W, Howarth, I.D., Harries, T.J.: 2005, \journal{Mon. Not. R. Astron. Soc.}, \vol{357}, 304.

Mao, S.A., Gaensler, B.M., Stanimirovi{\'c}, S., Haverkorn, M., McClure-Griffiths, N.M., Staveley-Smith, L., Dickey, J.M.: 2008, \journal{Astrophys. J.}, \vol{688}, 1029.

Matthews, D., Staveley-Smith, L., Dyson, P., Muller, E.: 2009, \journal{Astrophys. J. Let.}, \vol{691}, L115.

Payne, J.L., Filipovi{\'c}, M.D., Reid, W., Jones, P.A., Staveley-Smith, L., White, G.L.: 2004, \journal{Mon. Not. R. Astron. Soc.}, \vol{355}, 44.

Sault, R.J., Killeen, N.: 2010, Miriad Users Guide, ATNF, Sydney.

Sault, R.J., Wieringa, M.H.: 1994, \journal{Astron. Astrophys. Suppl. Series}, \vol{108}, 585.

Stanimirovi\'c, S.: 2002, \journal{Astron. Soc. of the Pacific Conf. Ser.}, \vol{278}, 375.

Staveley-Smith,~L., Sault, R.J., Hatzidimitriou, D., Kesteven, M.J.; McConnell, D.: 1997, \journal{Mon. Not. R. Astron. Soc.}, \vol{289}, 280.

Steer, D.G., Dewdney, P.E., Ito, M.R.: 1984, \journal{Astron. Astrophys.}, \vol{137}, 159.

\endreferences

}
\end{multicols}

\vfill\eject

{\ }



\naslov{NOVO PROUQAVA{NJ}E MALOG MAGELANOVOG OBLAKA U RADIO-KONTINUMU NA 20~CM: DEO~{\bf I} - SNIMCI}


\authors{G. F. Wong$^{\bf 1}$, M. D. Filipovi\'c$^{\bf 1}$, E. J. Crawford$^{\bf 1}$ A. De Horta$^{\bf 1}$,
T. Galvin$^{\bf 1}$,} 
\authors{D. Dra\v skovi\'c$^{\bf 1}$, J. L. Payne$^{\bf 1}$ }

\vskip3mm


\address{$^1$School of Computing and Mathematics, University of Western 
Sydney\break Locked Bag 1797, Penrith South DC, NSW 1797, Australia}

\Email{m.filipovic}{uws.edu.au}

\vskip3mm


\centerline{\rrm UDK \udc}

\vskip1mm

\centerline{\rit Originalni nauqni rad}

\vskip.7cm

\begin{multicols}{2}

{


\rrm

U ovoj studiji predstav{lj}amo nove {\rm ATCA} rezultate posmatra{nj}a visoke rezolucije i oset{lj}ivosti u radio-kontinumu Malog Magelanovog Oblaka (MMO) na {\rm $\lambda$=20~cm ($\nu$=1.4~GHz)}. Nove radio-mape nastale su spaja{nj}em arhivskih mozaik posmatra{nj}a na 20~cm sa Australija Teleskop Online Arhiva i podataka sa Parks radio-teleskopa (otvor antene 64~m). Naxi novi snimci su rezolucije od $\sim$10\arcsec\ do $\sim$150\arcsec\ i oset{lj}ivosti od {\rm r.m.s.=$\geq$0.5~mJy/beam}. Ovi snimci {\cc}e biti korix{\cc}eni u budu{\cc}im istra{\zz}iva{nj}ima kako objekata tako i ukupne strukture \mbox{MMO-a}.

}

\end{multicols}

\end{document}